# Measurements of the quantum geometric tensor in solids

Mingu Kang[1,2*], Sunje Kim[3,4,5*], Yuting Qian[3,4,5], Paul M. Neves[1], Linda Ye[1‡], Junseo Jung[3,4,5], Denny Puntel[6], Federico Mazzola[7,8], Shiang Fang[1], Chris Jozwiak[9], Aaron Bostwick[9], Eli Rotenberg[9], Jun Fuji[7], Ivana Vobornik[7], Jae-Hoon Park[2,10], Joseph G. Checkelsky[1], Bohm-Jung Yang[3,4,5†] & Riccardo Comin[1†]

[1]Department of Physics, Massachusetts Institute of Technology; Cambridge, Massachusetts 02139, USA.
[2]Max Planck POSTECH/Korea Research Initiative, Center for Complex Phase of Materials; Pohang 790-784, Republic of Korea.
[3]Department of Physics and Astronomy, Seoul National University; Seoul 08826, Republic of Republic of Korea.
[4]Center for Theoretical Physics (CTP), Seoul National University; Seoul 08826, Republic of Republic of Korea.
[5]Institute of Applied Physics, Seoul National University; Seoul 08826, Republic of Korea.
[6]Department of Physics, University of Trieste; Trieste 34127, Italy.
[7]Istituto Officina dei Materiali, Consiglio Nazionale delle Ricerche; Trieste 34149, Italy.
[8]Department of Molecular Sciences and Nanosystems, Ca' Foscari University of Venice; 30172 Venice, Italy.
[9]Advanced Light Source, E. O. Lawrence Berkeley National Laboratory; Berkeley, California 94720, USA.
[10]Department of Physics, Pohang University of Science and Technology; Pohang 790-784, Republic of Korea.

*These authors contributed equally to this work.

‡Current address: Department of Applied Physics, Stanford University; Stanford, California 94305, USA.

†Corresponding author. Email: rcomin@mit.edu; bjyang@snu.ac.kr

**Understanding the geometric properties of quantum states and their implications in fundamental physical phenomena is at the core of modern physics[1]. The Quantum Geometric Tensor (QGT) is a central physical object in this regard, encoding complete information about the geometry of the quantum state. The imaginary part of the QGT is the well-known Berry curvature, which plays a fundamental role in the topological magnetoelectric and optoelectronic phenomena. The real part of the QGT is the quantum metric, whose importance has come to prominence very recently, giving rise to a new set of quantum geometric phenomena, such as anomalous Landau levels[2,3], flat band superfluidity[4,5], excitonic Lamb shifts[6], and nonlinear Hall effect.[7] Despite the central importance of the QGT, its experimental measurements have been restricted only to artificial two-level systems[8–10]. In this work, we develop a framework to measure the QGT (both quantum metric and Berry curvature) in crystalline solids using polarization-, spin-, and angle-resolved photoemission spectroscopy. Using this framework, we demonstrate the effective reconstruction of the QGT in solids in the archetype kagome metal CoSn, which hosts topological flat bands. The key idea is to introduce another geometrical tensor, the quasi-QGT, whose components, the band Drude weight and orbital angular momentum, are experimentally accessible and can be used for extracting the QGT. Establishing such a momentum- and energy-resolved spectroscopic probe of the QGT is poised to significantly advance our understanding of quantum geometric responses in a wide range of crystalline systems.**

The quantum geometric tensor (QGT) is a crucial entity that characterizes the geometric properties of quantum states, playing a significant role in various fundamental physical phenomena.[1] In general, the QGT is a complex quantity comprised of the quantum metric and the Berry curvature as its real and imaginary parts, respectively (Fig. 1c,d)[11,12]. The Berry curvature describes the fictitious magnetic field in the parameter space of a particle under adiabatic motion (Fig. 1c). The anomalous velocity driven by the Berry curvature underlies various topological and quantized phenomena including the canonical Aharonov-Bohm effect, the quantum anomalous Hall effect and quantized Faraday/Kerr rotation in magnetic topological insulators[13–16], the valley Hall effect in monolayer transition metal dichalcogenides[17], and nonlinear optical responses, e.g., the circular photogalvanic effect at the surface of topological insulators[18]. The counterpart of the Berry curvature – the quantum metric – describes the distance between quantum states in the parameter space (Fig. 1d), which is related to quantum fluctuations[19,20], dissipative responses[21], and fidelity susceptibility[22] of the system. Over the past few years, it has been increasingly recognized that the quantum metric can give rise to a new set of quantum geometric phenomena distinct from well-known Berry curvature effects, including anomalous Landau levels[2,3], flat band superfluidity[4], geometric orbital susceptibility[23,24], and excitonic Lamb-shifts[6]. The observable consequences of the underlying quantum metric have become particularly apparent in recent experimental reports of quantum metric effects on Moiré flat band superconductivity[5] and on a new type of quantum geometrical Hall effect[7]. These studies altogether highlight the central role played by the QGT in increasingly diverse range of geometric phenomena in solids.

Despite its central importance, the experimental measurement of the QGT has mainly been reported in engineered two-level systems, such as superconducting qubits[8], diamond NV centers[9], and planar microcavities with embedded quantum wells[10]. On the other hand, the direct measurements of the QGT in solids, as a function of electron energy and momentum, has remained a challenging task. Given that the electronic band structure in solids typically comprises numerous bands, novel strategies for quantifying quantum geometric effects in solids necessitate new techniques for extracting the QGT that are broadly applicable to multi-band systems.

Here, we introduce the procedure to experimentally extract the QGT (both the Berry curvature (BC) and the quantum metric (QM)) of Bloch electrons in periodic crystals. Central to our study is the introduction of new geometric tensor, the quasi-QGT, whose real and imaginary parts correspond to the band Drude weight (BDW) and the orbital angular momentum (OAM), respectively (Fig. 1a,b). The quasi-QGT is a faithful representation of the QGT such that it is proportional to the QGT in two-band systems while it is an excellent approximation in multi-band systems. The clear physical meaning of the quasi-QGT allows its direct measurement using photoemission spectroscopy to ultimately extract the QGT of Bloch electrons. To demonstrate the idea, we investigate the prototypical kagome metal CoSn as a test case[25], and present the experimental reconstruction of quantum metric and spin Berry curvature which contain the direct signatures of topological flat bands. The broad applicability of our approach is further demonstrated using other model two-dimensional materials, including hexagonal boron nitrides and monolayer black phosphorous (Supplementary Discussion 5). The framework developed here to measure the QGT in solids serves as a stepping stone towards uncovering and engineering the wider QGT phenomena in diverse materials with complex internal structures.

**Outline of the QGT measurements in solid**

Before presenting the experimental measurement, we describe our theoretical idea to measure the QGT with the overall strategy illustrated in Fig. 1. The QGT $Q_{ij}^n(\mathbf{k})$ of a non-degenerate Bloch eigenstate $|\psi_n(\mathbf{k})\rangle$ with crystal momentum $\mathbf{k}$ and band index $n$ is given by[22]

$$Q_{ij}^n(\mathbf{k}) = \langle \partial_{k_i}\psi_n(\mathbf{k})|1 - P_n(\mathbf{k})|\partial_{k_j}\psi_n(\mathbf{k})\rangle, \tag{1}$$

where $\partial_{k_i} = \partial/\partial k_i$ and $P_n(\mathbf{k}) = |\psi_n(\mathbf{k})\rangle\langle\psi_n(\mathbf{k})|$ is the projection operator onto $|\psi_n(\mathbf{k})\rangle$. The real and imaginary parts of the QGT $Q_{ij}^n(\mathbf{k})$ encode the QM $G_{ij}^n(\mathbf{k})$ and BC $F_{ij}^n(\mathbf{k})$ (Fig. 1c,d):

$$G_{ij}^n(\mathbf{k}) = \mathrm{Re}[Q_{ij}^n(\mathbf{k})] = \sum_{m\neq n} \mathrm{Re}\left[\frac{\langle\psi_n|\partial_{k_i}H(\mathbf{k})|\psi_m\rangle\langle\psi_m|\partial_{k_j}H(\mathbf{k})|\psi_n\rangle}{(E_m - E_n)^2}\right], \tag{2}$$

$$F_{ij}^n(\mathbf{k}) = -2\mathrm{Im}[Q_{ij}^n(\mathbf{k})] = -2\sum_{m\neq n} \mathrm{Im}\left[\frac{\langle\psi_n|\partial_{k_i}H(\mathbf{k})|\psi_m\rangle\langle\psi_m|\partial_{k_j}H(\mathbf{k})|\psi_n\rangle}{(E_m - E_n)^2}\right], \tag{3}$$

where $H(\mathbf{k})$ is the Bloch Hamiltonian and $E_n(\mathbf{k})$ is the energy of $|\psi_n(\mathbf{k})\rangle$. To probe the momentum space distribution of QM and BC with spectroscopic techniques such as angle-resolved photoemission spectroscopy (ARPES), it is essential to find their relation to experimentally measurable momentum-dependent quantities. We resolve this issue by introducing another geometric quantity $q_{ij}^n(\mathbf{k})$, dubbed the quasi-QGT, defined as

$$q_{ij}^n(\mathbf{k}) = \langle\partial_{k_i}\psi_n(\mathbf{k})|H(\mathbf{k}) - E_n(\mathbf{k})|\partial_{k_j}\psi_n(\mathbf{k})\rangle. \tag{4}$$

The real and imaginary parts of the quasi-QGT are given by:

$$g_{ij}^n(\bm{k}) = \text{Re}[q_{ij}^n(\bm{k})] = \sum_{m \neq n} \text{Re}\left[\frac{\langle\psi_n|\partial_{k_i}H(\bm{k})|\psi_m\rangle\langle\psi_m|\partial_{k_j}H(\bm{k})|\psi_n\rangle}{E_m - E_n}\right], \quad (5)$$

$$f_{ij}^n(\bm{k}) = \text{Im}[q_{ij}^n(\bm{k})] = \sum_{m \neq n} \text{Im}\left[\frac{\langle\psi_n|\partial_{k_i}H(\bm{k})|\psi_m\rangle\langle\psi_m|\partial_{k_j}H(\bm{k})|\psi_n\rangle}{E_m - E_n}\right]. \quad (6)$$

Both the real and imaginary parts of the quasi-QGT hold well-defined physical meaning. The imaginary part $f_{ij}^n(\bm{k})$ is nothing but the OAM of the wave packet made of $|\psi_n(\bm{k})\rangle$[26,27] (Fig. 1b). At the same time, the real part $g_{ij}^n(\bm{k})$ is related to the effective band curvature of $|\psi_n(\bm{k})\rangle$ (Fig. 1c, see below for details) and represents the band contribution to the momentum-resolved Drude weights or band Drude weight (BDW) in metallic systems[28,29]. The quasi-QGT with well-defined physical meanings (BDW and OAM) complements the QGT (QM and BC) for the description of the intrinsic geometric properties of the Bloch electrons.

The similar structure of the QGT in Eq. (**2**), (3) and the quasi-QGT in Eq. (5), (6) suggests that the elusive QGT can be determined through the measurement of the quasi-QGT, which is experimentally accessible. In two-band systems, the QGT and the quasi-QGT are proportional to each other (i.e. $Q_{ij} = q_{ij}/\Delta E$, where $\Delta E$ is the energy difference between the two bands), thus the momentum space distribution of the QGT can be unambiguously determined by measuring the quasi-QGT. In real materials with multiple bands, a similar relation holds for pairs of bands which are energetically well separated from or not hybridized with other bands at a given momentum (i.e. $Q_{ij} \approx q_{ij}/\Delta E$).

We now elaborate on how the components of the quasi-QGT, namely the BDW and OAM, can be accessed experimentally. First, the OAM distribution in momentum space can be estimated by circular dichroism (CD)-ARPES experiments[30–33], in which one measures the difference in photoemitted electron intensity using right-handed and left-handed circularly polarized light carrying $\pm 1$ angular momentum, respectively. As detailed in the Supplementary Discussion 6 and Ref.[32], CD-ARPES predominantly measures the difference between $m = \pm 1$ OAM components of the gauge-transformed wave packet where $m$ indicates the quantum number of the $z$-component of the angular momentum. Therefore, the OAM and the CD-ARPES signal ($I_{CR} - I_{CL}$) exhibit similar textures when the angular momentum $\pm 1$ states are dominant. We note that the angular momentum of the quantum state is strongly influenced by the phase difference between its orbital wave functions components as well as the atomic orbital characters as explained in detail in Supplementary Discussion 7.

Next, the BDW can be estimated by the band curvature analysis. To illustrate this, we take the derivative $\partial_{k_i}\partial_{k_j}$ on both sides of the equation $E_n = \langle \psi_n|H|\psi_n \rangle$, which gives[28]

$$g_{ij}^n(\boldsymbol{k}) = \frac{1}{2}\left(\langle \psi_n|\partial_{k_i}\partial_{k_j}H|\psi_n\rangle - \partial_{k_i}\partial_{k_j}E_n\right). \tag{7}$$

Here, the second term on the right-hand side is nothing but the curvature of the band $E_n(k)$ which can be directly measured by ARPES, while the physical meaning of the first term is not transparent. However, interestingly, we find that in many cases – including the case when the nearest-neighbor (NN) hopping is dominant and also the case when the difference between NN and next-NN distances are not significant (Supplementary Discussion 4,5) – the trace of the BDW can be approximated as

$$g_{xx}^n + g_{yy}^n \approx \tilde{g}^n = \frac{1}{2}\left(-a^2(E_n - E_0) - \left(\partial_{k_x}^2 + \partial_{k_y}^2\right)E_n\right), \tag{8}$$

where $a$ is the distance between NN sites, and $E_0$ is the onsite potential of $|\psi_n\rangle$. Since $\tilde{g}^n$ is related to the band curvature $\left(\partial_{k_x}^2 + \partial_{k_y}^2\right)E_n$ and approximates $g^n$, we call $\tilde{g}^n$ as the effective BDW. All the terms in Eq. (8) can be obtained from band structure data. This indicates that both real and imaginary parts of the quasi-QGT (BDW and OAM) can be extracted from the analysis of the ARPES data (band curvature and CD), which ultimately enables the estimation of QGT (QM and BC). This overall procedure to obtain QGT in solid is summarized in Fig. 1.

**QGT in the kagome lattice toy model**

To demonstrate our approach, we consider a simple tight-binding model on the kagome lattice with an *s*-orbital per site and the Kane-Mele type spin-orbit coupling (SOC) (see also Method and Supplemental Discussion 2). The SOC lifts the degeneracy between the two Dirac bands at K and at the quadratic band touching between the flat band and the parabolic band at Γ, as shown in Fig. 2c. In this model, because the inversion ($P$) and time-reversal ($T$) symmetries are simultaneously preserved, each band is doubly degenerate and thus the QGT should take a non-abelian form[22]. However, due to the $M_z$ mirror symmetry about the two-dimensional kagome plane ($M_z = i\sigma_z$, where $\sigma_z$ is the *z*-directional spin operator), each spin sector is decoupled; therefore, the abelian form of the QGT and the quasi-QGT introduced above can be applied to each spin channel. For the up-spin and down-spin channels, we obtain the same QM and BDW, but oppositely signed BC and OAM with the same magnitude (see Supplementary Discussion 1).

Fig. 2d,e,h, and i show the $g$ (BDW), $G$ (QM), $\tilde{g}$ (our estimation of BDW using Eq.(8)) and $\tilde{g}/\Delta E$ (our estimation of the QM under the two-band approximation) of the lower Dirac band in the up-spin channel around Γ point, respectively. In this case, $\Delta E$ is the energy difference between the flat band and the lower Dirac band: $\Delta E = E_{flat} - E_{Dirac}$. To compute $\tilde{g}$, $E_0$ in Eq. (8) is estimated by the energy of the lower Dirac band at the momentum M (see Supplementary Discussion 4). One can clearly observe that $g$ and $\tilde{g}$ have almost the same texture. Their tiny difference in magnitude comes from the Kane-Mele SOC neglected in the approximation $g_{xx} + g_{yy} \approx \tilde{g}$. As shown in Fig. 2e and i, QM and $\tilde{g}/\Delta E$ are also nearly identical due to the small energy gap between the lower Dirac band and the nearly flat band at Γ.

In Fig. 2f,g,j, and k, we compare the $f$ (OAM), $F$ (BC), the normalized CD-ARPES signal $I_{CD}^{norm} = (I_{CR} - I_{CL})/(I_{CR} + I_{CL})$, and $I_{CD}^{norm}/\Delta E$, respectively. The photoemission intensity is obtained by using the Fermi golden rule in the dipole approximation (see Methods). As shown in Fig. 2f, a nonzero OAM appears around the SOC gap at Γ, reflecting orbital loop currents around the center of the kagome unit cell. The normalized CD-ARPES signal in Fig. 2j correctly captures the dominant OAM signals around Γ, but misses the dip structure at Γ. Consistent with recent reports[30,31,33], the texture of the CD-ARPES signal and OAM is similar, but some discrepancy remains. In Supplementary Discussion 6,7, we reveal the microscopic origin of the difference between the CD-ARPES and OAM in a specific momentum range and attribute the discrepancy in our model to the sizable contributions from the $|m_l| \geq 2$ angular momentum states near Γ. The BC and the normalized CD-ARPES signal divided by the energy difference exhibit a similar tendency as shown in Fig. 2g and k. To sum up, we find that the quasi-QGT and QGT are well estimated by the effective BDW $\tilde{g}$ and the CD-ARPES, both of which are experimentally accessible. In Supplementary Discussion 5, we demonstrated the broad applicability of our procedure to other 2D systems, such as hexagonal boron nitrides and monolayer black phosphorous. In Supplementary Discussion 8, we also discuss how our procedure can be extended to more general cases, where more than two bands become vicinal to each other in different parts of the Brillouin zone. The reconstructed quantum metric obtained through this extended method faithfully reflects the positions of the maxima and the minima of the real quantum metric, which contain information about the rate of change of the quantum state in momentum space.

**Measurement of the quantum metric from the band curvature analysis**

We applied the procedure developed above to the prototypical kagome metal cobalt tin (CoSn). CoSn belongs to the class of binary intermetallics $T_mX_n$[34], which recently attracted great interest as a platform to realize quantum matter phenomena including Chern insulators, magnetic Weyl semimetals, and topological flat bands[25,35,36]. CoSn is an ideal candidate for momentum- and energy-resolved studies of quantum geometric tensor. CoSn forms a simple lattice structure with highly two-dimensional kagome layers (Fig. 3a,b)[25,34], and exhibits negligible magnetic and electronic correlations, meaning its band structure is well captured by a noninteracting model[37,38]. Further, the decoupling between the $3d$ orbital degrees of freedom validates the adoption of the two-band approximation (see Supplementary Discussion 4). As shown in the ARPES spectra of Fig. 3c, the low-energy electronic structure of CoSn is a clean realization of the canonical band structure of the kagome lattice (Fig. 3e), including the characteristic flat band and the quadratic band bottom from the Dirac band[25,39]. The experimental dispersion of CoSn is closely reproduced by the DFT band calculation (Fig. 3d), revealing that both the flat and the Dirac band originate from Co $d_{xz}$ orbitals. The ARPES data in Fig. 3c also demonstrate that the band touching degeneracy between the Dirac and flat band at the zone center (Γ point) is lifted in CoSn by the intrinsic SOC from the $3d$ electrons. This results in a prominent SOC gap at Γ ($\Delta_{SOC} \approx 76 \pm 5$ meV) and endows the nontrivial topology to the Dirac and flat bands[25]. Below, we demonstrate how the quantum metric and spin Berry curvature can be probed experimentally near the avoided band touching point at Γ with a topological gap.

Fig. 3f displays the high-resolution ARPES spectrum of CoSn measured under the experimental geometry designed to maximize the photoelectron yield from the $d_{xz}$ Dirac bands while suppressing the signal from the other bands (see Methods). The optimization of the photoemission matrix elements allows us to isolate the band of interest and to determine its dispersion and curvature reliably. What is already apparent from the raw data is a flattening of the Dirac band bottom near the Γ point, signaling a modification of the dispersion induced by the opening of the SOC gap. To quantify this effect, we fit the Dirac band dispersion to a kagome tight-binding model with and without SOC (Fig. 3g). The experimental dispersion is closely captured by the model with SOC, while the one without SOC deviates especially near the band bottom. This band renormalization near the topological gap is reflected in the curvature of the dispersion (proportional to the band effective mass $m^*$): as shown in the upper panel of Fig. 3g, the SOC dramatically suppresses the curvature near $\Gamma$ and induces a nontrivial momentum dependence of $m^*$ (black solid line), while the curvature remains relatively featureless in the dispersion without SOC (grey dashed line). This renormalization of the band curvature, observed in our experimental dispersion, encodes the quantum geometric information of the Bloch states (Eq. (8) above) and allows us to extract the BDW and QM as outlined in Figs. 1, 2.

We extend the analysis to the two-dimensional momentum space and reconstruct the BDW and QM in CoSn. For this, we track the two-dimensional dispersion of the Dirac band $E_{Dirac}(k_x, k_y)$ (Fig. 3h) by fitting an array of ARPES energy distribution curves from a 21×21 $k_x$-$k_y$ grid in the momentum range –0.26 ~ 0.26 Å$^{-1}$ (Supplementary Fig. S1). For the curvature calculation, we first fit $E_{Dirac}(k_x, k_y)$ to a sixth-order polynomial as shown in Fig. 3i ($E^{fit}_{Dirac}(k_x, k_y)$). The use of a generic polynomial provides an effective and unbiased approach to remove small noise level in the raw $E_{Dirac}(k_x, k_y)$ data (which can create significant artifacts in the second derivative) while preserving the intrinsic curvature of the experimental dispersion $E_{Dirac}(k_x, k_y)$. The curvature obtained from $E^{fit}_{Dirac}(k_x, k_y)$ displays a strong renormalization and suppression near the SOC gap, in accordance with the results in Fig. 3g (Supplementary Fig. S2). The experimentally obtained $\tilde{g} = \frac{1}{2}(-a^2(E_{Dirac} + E_0) - (\partial^2_{k_x} + \partial^2_{k_y})E^{fit}_{Dirac})$, which approximates the BDW, and $\tilde{g}/\Delta E = \tilde{g}/(E_{flat} - E_{Dirac})$, which approximates the QM, are shown in Fig. 3j,k, respectively. For comparison, we also present in Fig. 3l,m the theoretical BDW and QM obtained from the realistic $\{d_{xz}, d_{yz}\}$ orbital-based kagome tight-binding model developed for CoSn (see Supplementary Discussion 3). The close correspondence between the experiment and theory validates the method of estimating the BDW and QM using $\tilde{g}$ and $\tilde{g}/\Delta E$, and establishes a new way to measure the QM in crystalline materials. Although we calculated the quantum metric of the Dirac band, it is nearly the same as the quantum metric of the flat band near the SOC-driven band gap. This implies that our reconstructed quantum metric provides direct information on the topology of the flat band, which is further explained in Supplementary Discussion 10.

**Measurement of the Berry curvature using spin-resolved CD-ARPES**

For a complete reconstruction of the quasi-QGT and QGT, one needs to probe their respective imaginary parts – the OAM and Berry curvature (Fig. 1). As demonstrated in other systems, including transition metal dichalcogenides[30,40], topological insulators[41,42] and strong spin-orbit coupled materials[43,44], the momentum-resolved OAM and BC can be extracted using CD-ARPES, though the exact correspondence between CD-ARPES and OAM is highly dependent on the details of materials and microscopic models. Notably, all previous CD-ARPES studies of OAM and BC have been limited to systems with reduced symmetry, where the breaking of local inversion symmetry in bulk or at the surface produces a nonvanishing and measurable OAM and Berry curvature. In the case of CoSn, both inversion and time-reversal symmetries are preserved so that the OAM and Berry curvature vanish everywhere in the Brillouin zone. Even in this case, however, the *hidden* OAM and Berry curvature in each spin sector – relevant for spin Chern insulators or topological insulators – are expected to be nonzero and concentrated near the SOC gap. Especially, the out-of-kagome-plane spin $S_z$ remains a good quantum number in CoSn even in the presence of SOC, owing to the quasi two-dimensionality of the electronic structure. We thus designed a spin-resolved ARPES measurement of the kagome bands in CoSn and obtained the CD-ARPES signal projected to each spin sector (Fig. 4a). Our experiment is in part motivated by the recent theoretical proposal for the use of spin-resolved CD-ARPES to detect the hidden Berry curvature in two-dimensional materials[32].

Fig. 4b,c display the spin-integrated ARPES spectrum and energy distribution curve at Γ, where the photoemission intensity from the circular right ($I_{CR}$) and circular left ($I_{CL}$) incoming light polarization have been summed up (the experimental geometry is shown in Fig. 4a). This geometry exclusively highlights the photoemission from the bands of interest, i.e., the $d_{xz}$ Dirac and flat bands with SOC gap in between. The corresponding spin-resolved CD-ARPES signal on the up-spin channel $I_{CD}(\uparrow_z)$ is shown in Fig. 4d,e (see also Methods and the Fig. 4 caption for how we obtained the CD). A clear signature of circular dichroism is found across the SOC gap, decaying further away from Γ. Note that extrinsic contributions to the dichroism, such as geometric CD[45], and photoelectron final state effects[46,47], can be ruled out based on the experimental geometry used and the anisotropy of the CD-signal (see below). The observed spin-resolved CD-ARPES signal thus reflects the intrinsic geometric properties of the Bloch wavefunctions near Γ. As shown in Fig. 4f,g, the CD-ARPES signal for the down-spin channel $I_{CD}(\downarrow_z)$ exhibits similar momentum

dependence but the opposite sign of CD. This observation is consistent with the time-reversal symmetry of CoSn and supports the intrinsic origin of the observed signal. Finally, Fig. 4h,i show that the CD-ARPES signal vanishes for the in-plane spin channel $\uparrow_x$, confirming the dominance of the orbital angular momentum in the out-of-plane direction, as expected from the two-dimensional nature of this system (see also Supplementary Fig. S3 and Supplementary Discussion 9 for the photon-energy dependent spin CD-ARPES measurements).

The observed behavior of the CD-ARPES signal – concentrated near the SOC gap, with sign change under the time-reversal symmetry, and out-of-plane polarization – reflects the properties of the underlying OAM and BC. As shown in Fig. 4l-n, the OAM and BC of the Dirac band calculated from the $\{d_{xz}, d_{yz}\}$ orbital-based kagome tight-binding model are also distributed around the SOC gap (panels l,m), their signs are flipped upon spin reversal (panels l,m), and they vanish for in-plane spin projections (panel n). This demonstrates the intimate connection between the spin-resolved CD-ARPES and hidden OAM and BC in CoSn. For a more quantitative comparison, Fig. 4j displays the theoretical OAM and the normalized CD-ARPES intensity $I_{CD}^{norm}(\uparrow_z) = (I_{CR}(\uparrow_z) - I_{CL}(\uparrow_z))/(I_{CR}(\uparrow_z) + I_{CL}(\uparrow_z))$ of the Dirac band along the $\Gamma$-M momentum-space direction. A comparison between the BC and $-2I_{CD}^{norm}(\uparrow_z)/(E_{flat} - E_{Dirac})$ is presented in Fig. 4k. The experimental CD-ARPES signal correctly captures the decay of the OAM and BC toward the zone boundary M. However, there is a noticeable difference at the zone center: the CD-ARPES intensity is maximized near $\Gamma$, while the OAM and BC vanish. The latter can be understood to arise from the site-centered inversion symmetry of the kagome lattice (see Supplementary Discussion 4). In general, a mismatch between OAM and CD-ARPES in certain momentum ranges is expected to some degree since the CD-ARPES predominantly measures the orbital polarization of the wave function projected to the $m_l = \pm 1$ state, while all $m_l$ states contribute to the total OAM[32,40]. As detailed in the Supplementary Discussion 7, the sizable contribution from the $|m_l| \geq 2$ angular momentum states is indeed revealed in our model near the $\Gamma$ point, which arises from the nontrivial $C_3$ rotational eigenvalues of the kagome bands. Based on this understanding, we also specified the conditions where such a discrepancy between the CD-ARPES and OAM can vanish (Supplementary Discussion 7).

To conclude, we report the protocol to experimentally extract the QGT in a crystalline solid. Using the highly symmetric kagome metal CoSn with strongly two-dimensional band structure, we have unveiled and characterized the elusive spin-BC and its conjugate QM via polarization-,

spin-, and angle-resolved photoemission experiments, thus establishing the nontrivial band geometry of a spin-orbit coupled crystal. Our strategy to measure the QGT is generally applicable to any crystalline solid, and the obtained QGT information can be effectively used to determine the nontrivial characteristics of the bands, such as their nonlinear response [7] and Chern numbers (Supplementary Discussion 10). The spectrum of quantum geometric phenomena is continuously and rapidly expanding – the new methodology presented here is poised to provide a complete characterization of quantum geometry in a wide range of material systems, ultimately to offer advanced understanding of diverse quantum geometric phenomena.

## Methods

**Sample synthesis**

Single crystals of CoSn were synthesized using a Sn self-flux method. Cobalt powder (Alfa Aesar, 99.998%) and tin pieces (Alfa Aesar, 99.9999%) were mixed with a molar ratio of 1:9 and loaded in an alumina crucible and sealed in a quartz tube under high vacuum. The tube was heated to 950 °C and maintained for 5 h, and cooled gradually to 650 °C with a rate of 2–3 °C/h. At 650 °C the tube was removed from the furnace. Crystals were separated from the Sn flux by the centrifuge dissociation procedure. CoSn crystalizes in a hexagonal prismatic shape with typical dimension 1×1×5 ~ 2×2×10 mm³ (longest dimension along [001]).

**High-resolution ARPES measurements**

High-resolution ARPES data presented in Fig. 3 were obtained at beamline 7.0.2 (MAESTRO) of the Advanced Light Source equipped with a R4000 hemispherical analyzer (Scienta Omicron) and a custom-designed slit deflector. For the ARPES experiments, the atomically flat surfaces of CoSn are prepared in two different ways, one by fine polishing followed by *in situ* Ar ion sputtering and annealing above 800 ºC, and the other by fracturing the CoSn rods inside the ultrahigh vacuum better than $4\times10^{11}$ torr. We carefully optimized the experimental geometry and photoemission matrix elements to enhance the photoemission from the bands of interest. The data in Fig. 3c (3f) was measured with LV polarized 70 eV (128 eV) photons and with the sample Γ-M direction aligned parallel (perpendicular) to the scattering plane. Both 70 eV and 128 eV photons measure the $k_z = 0$ plane of the CoSn Brillouin zone. All experiments were performed at 7 K. Throughout the text and figures, we consistently defined the Γ-K high symmetry direction as $k_x$ and the Γ-M direction as $k_y$.

**Spin CD-ARPES measurements**

Spin CD-ARPES data presented in Fig. 4 were obtained at the APE-LE beamline of Elettra synchrotron. The endstation is equipped with a DA30 electron analyzer (Scienta Omicron) and double VLEED spin detector allowing the simultaneous measurements of in-plane ($x$, $y$) and out-of-plane ($z$) spin polarizations (see Fig. 4a)[48,49]. To avoid an artifact from the polarization-dependent beam intensity, we calculated $I_{CD}(\uparrow_z)$ as the difference between $I_{CR}(\uparrow_z)$ and $I_{CR}(\downarrow_z)$ (i.e., $I_{CD}(\uparrow_z) = I_{CR}(\uparrow_z) - I_{CR}(\downarrow_z)$) using the fact that $I_{CR}(\downarrow_z)$ is identical to the $I_{CL}(\uparrow_z)$ by the time-reversal symmetry. Similarly, $I_{CD}(\downarrow_z)$ is obtained as $I_{CD}(\downarrow_z) = I_{CL}(\uparrow_z) - I_{CL}(\downarrow_z)$. All data in Fig. 4 are obtained with 70 eV photons. To rule out the photoelectron final state effect as an origin of the observed CD, we repeated the measurements with 76 eV and 83 eV photon energies and obtained qualitatively consistent results (see Supplementary Fig. S3). All measurements were performed at 78 K.

**Tight-binding model**

In the tight-binding approximation, we approximate the Bloch wavefunction $|\psi_{kn}\rangle$ by the Fourier transformation of the atomic orbitals $|w_j(\mathbf{R})\rangle$,

$$|\psi_{kn}\rangle = \frac{1}{\sqrt{N}} \sum_{\mathbf{R},j} e^{i\mathbf{k}\cdot(\mathbf{R}+\mathbf{x}_j)} C_j(\mathbf{k}) |w_j(\mathbf{R})\rangle,$$

where $\mathbf{k}$ is the crystal momentum, $\mathbf{R}$ is the lattice vector, $N$ is the number of unit cells, j is the index for the atomic orbitals and spin, and $\mathbf{x}_j$ is the sublattice position of $|w_j(\mathbf{R})\rangle$. To find the coefficients $C_j(\mathbf{k})$ of $|w_j(\mathbf{R})\rangle$ we solve the energy eigenvalue equation of the tight-binding Hamiltonian $H(\mathbf{k})$,

$$H(\mathbf{k})_{ij} = \sum_{\mathbf{R}} t_{ij}(\mathbf{R}) e^{-i\mathbf{k}\cdot(\mathbf{R}+\mathbf{x}_i-\mathbf{x}_j)},$$

where $t_{ij}(\mathbf{R})$ is the hopping or the SOC parameter. For the *s*-orbital kagome tight-binding model in Fig. 2, we consider nearest-neighbor (NN) hopping and the Kane-Mele type SOC. To make the $d_{xz}/d_{yz}$ kagome tight-binding model in Fig. 3, 4, we consider the NN and next NN hopping with onsite SOC and fit the model to the energy band data fitted from the CD-ARPES result, which has a similar band structure with the DFT calculation. The detailed tight-binding parameters we use and comparison among the model, DFT calculation, and the experimental data are attached in Supplementary Discussion 2,3.

**Theoretical calculation of CD-ARPES signal**

In Fig. 2g,h, we use $E_0$ as the onsite energy of the *s*-orbital. In Fig. 3j,k, we use $E_0$ as a constant which makes $\tilde{g} = 0$ at $\mathbf{k} = \Gamma$. The details are explained in Supplementary Discussion 4.

Using the tight-binding Hamiltonian, we compute the CD-APRES signal as follows. The photoemission intensity $I_{\hat{\varepsilon}}$ from the light with a polarization $\hat{\varepsilon}$ is proportional to the absolute square of the photoemission matrix element $M_n(\hat{\varepsilon}) = \langle \chi_p | \hat{\varepsilon} \cdot \mathbf{r} | \psi_{kn} \rangle$ under the dipole approximation, where $|\chi_p\rangle$ is the plane wave function with a momentum $\boldsymbol{p}$ and $|\psi_{kn}\rangle$ is the Bloch wave function with a crystal momentum $\boldsymbol{k}$. To calculate $M_n(\hat{\varepsilon})$ for the *s*-orbital kagome tight-binding model, we assume that the atomic orbitals $|w_j(\boldsymbol{R})\rangle$ are well localized. Also, the position $\boldsymbol{r}$ is measured relative to the center of a triangular unit cell. The details are explained in Supplementary Discussion 2.

**DFT calculations of Band structures**

Vienna Ab initio Simulation Package (VASP) package[50,51] with the projector augmented wave method[52,53] is employed to obtain electronic band structures for DFT calculations. The generalized gradient approximation (GGA) of Perdew-Burke-Ernzerhof (PBE) type[54] are used as the exchange-correlation functional. The energy cutoff for plane wave basis is set as 350 eV. We perform the Gamma centered 13× 13 × 10 k-mesh for self-consistent process. SOC is considered for band structure calculations. Crystal structure with $a = 5.308(6)$ Å and $c = 4.237(4)$ Å is obtained from our experiments.

**Acknowledgments**

This work was supported by the Air Force Office of Scientific Research grant FA9550-22-1-0432, and by the STC Center for Integrated Quantum Materials, NSF grant DMR-1231319. The work is funded in part by the Gordon and Betty Moore Foundation's EPiQS Initiative, Grant GBMF9070 to S.F., and J.G.C. M.K., and J.-H.P. were supported by National Research Foundation of Korea grant 2022M3H4A1A04074153. S.K., Y.Q., J.J., B.-J.Y. were supported by Samsung Science and Technology Foundation under Project Number SSTF-BA2002-06, and the National Research Foundation of Korea (NRF) grant funded by the Korea government (MSIT) No. 2021R1A5A1032996. P.M.N., J.G.C. were supported by ARO Grant No. W911NF-16-1-0034, and Center for Advancement of Topological Semimetals, an Energy Frontier Research Center funded by the US Department of Energy Office of Science, Office of Basic Energy Sciences, through the Ames Laboratory under Contract No. DE-AC02-07CH11358. L.Y. were supported by STC Center for Integrated Quantum Materials, NSF grant DMR-1231319, Heising-Simons Physics Research Fellow Program, and Tsinghua Education Foundation. D.P., F.M., J.F., I.V. were supported by Nanoscience foundry and fine analysis (NFFA-MUR Italy Progetti Internazionali) facility. This research used resources of the Advanced Light Source, a U.S. DOE Office of Science User Facility under contract no. DE-AC02-05CH11231. M.K. acknowledges a Samsung Scholarship from the Samsung Foundation of Culture.

**Author contributions**

M.K., S.K., B.-J.Y., and R.C. conceived the project; M.K., D.P., F.M., and R.C. performed the ARPES and spin-ARPES experiments and analyzed the resulting data with help from C.J., A.B., E.R., J.F., I.V., and J.-H.P; S.K. and Y.Q. performed the theoretical calculations with help from J.J. and S.F.; P.M.N. and L.Y. synthesized and characterized the crystals with supervision from J.G.C.; M.K., S.K., B.-J.Y., and R.C. wrote the manuscript with input from all authors.

**Data and materials availability**

The data and materials that support the findings of this study are available from the corresponding authors upon request.

**Competing interests**

Authors declare that they have no competing interests.

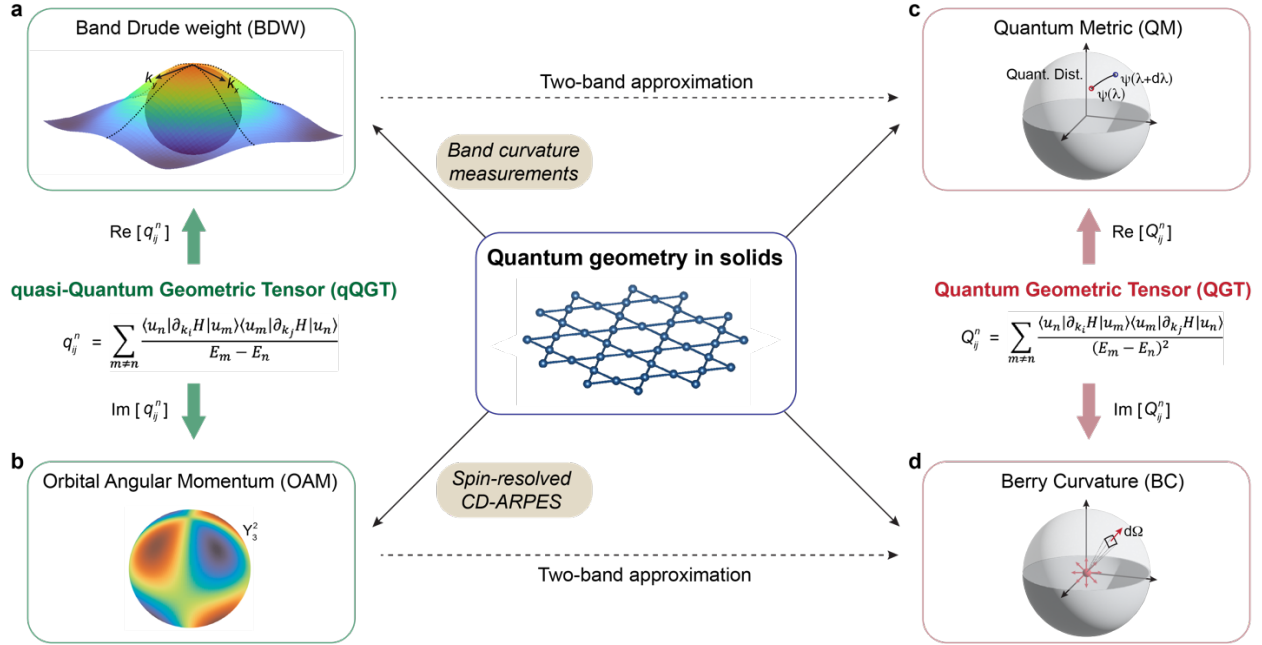

**Figure 1 | Strategy to measure the quantum geometric properties in condensed matter systems. a-d,** Here we focus on the two quantum geometric quantities, quasi-quantum geometric tensor (quasi-QGT) and quantum geometric tensor (QGT). A real part of the QGT is the quantum metric (QM), which measures the quantum distance between the states on the Bloch sphere (c). The imaginary part of the QGT is Berry curvature (BC), measuring the flux from the topological monopole at the origin (d). The real part of the quasi-QGT is the band Drude weight (BDW), i.e. band contribution to the momentum-resolved Drude weight (a), while its imaginary part corresponds to the orbital angular momentum (OAM) (b). The intrinsic connection between the quasi-QGT and QGT is evident from their expressions. For the experimental measurements of the QGT, we first estimated the BDW and OAM using band curvature and spin-resolved CD-ARPES measurements, respectively. Then we applied the two-band approximation to reconstruct the QM and BC, i.e. the full QGT.

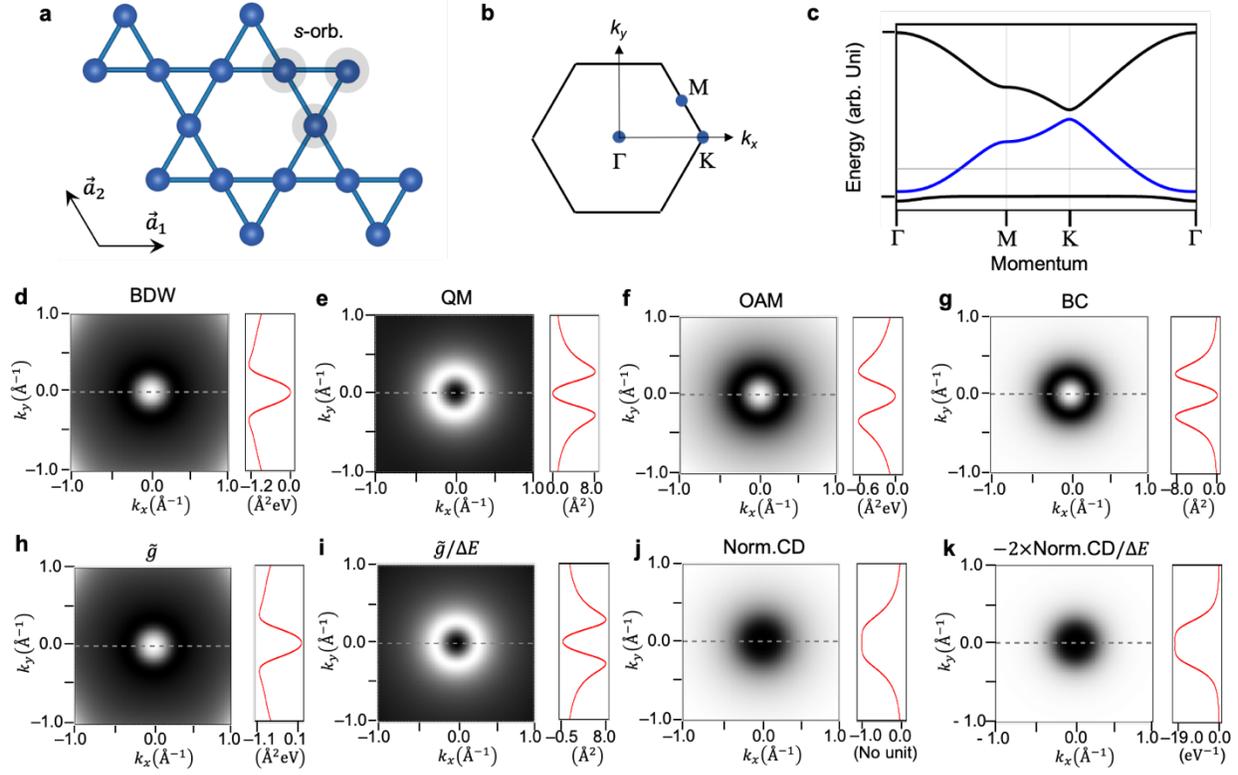

**Figure 2 | QGT, quasi-QGT, g̃, and CD-ARPES of the *s*-orbital kagome tight-binding model. a,b,** Lattice structure and corresponding Brillouin zone of the kagome lattice, respectively. **c,** Tight-binding band structure of the kagome lattice. Due to spin-orbit coupling, the lower Dirac band (blue solid line) is separated from the flat band near Γ and the upper Dirac band near K. **d-g,** Geometrical quantities of the wave function for the lower Dirac band in c. (d) BDW, (e)**,** QM, (f), OAM, (g), BC. **h-k,** Physical quantities of the lower Dirac band in c, measurable from the ARPES experiments. (h), g̃ (effective BDW), (i), g̃/$\Delta E$, (j), Normalized CD-ARPES intensity, (k), Normalized CD-ARPES intensity divided by $\Delta E$. Here $\Delta E = E_{flat} - E_{Dirac}$ is the energy difference between the flat band and the lower Dirac band.

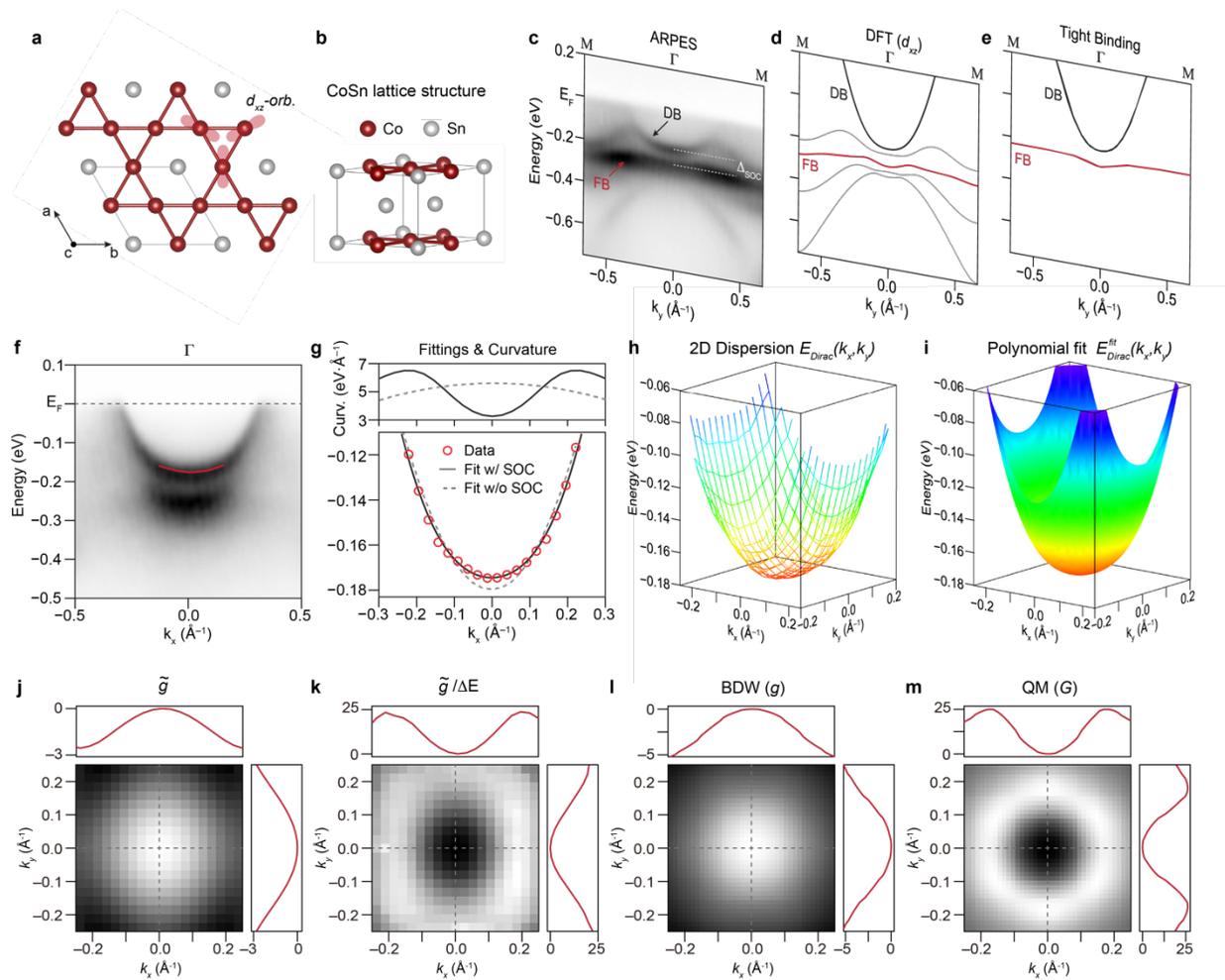

**Figure 3 | Experimental measurements of BDW and QM. a,b,** Top and side view of the crystal structure of CoSn. The neighboring Co kagome layers are spatially separated by the Sn spacer layer, realizing the quasi-two-dimensional kagome lattice dispersion in bulk CoSn. **c,** Experimental band structure of CoSn along the Γ-M high symmetry direction. Characteristic features of the kagome lattice dispersion, i.e., flat band, Dirac band, and SOC gap in between, are marked with the red, black, and white indicators, respectively. **d,** DFT band structure of CoSn. The Dirac and flat bands observed in C correspond to the kagome bands with $d_{xz}$ local orbital character, shown in black and red solid lines. **e,** Prototypical tight-binding band structure of the kagome lattice for comparison with c,d. **f,** High-resolution ARPES spectra of the $d_{xz}$ Dirac band. The red line is a guide to the eye, highlighting the band flattening near the SOC gap. **g,** Dispersion of the $d_{xz}$ Dirac band (red circles) and fittings to the kagome tight-binding models with and without SOC (black solid and grey dashed lines, respectively). Corresponding curvatures are plotted in the upper panel. **h,** Experimental dispersion of the $d_{xz}$ Dirac band in two-dimensional momentum-space, $E_{Dirac}(k_x, k_y)$. **i,** Fitting of the $E_{Dirac}(k_x, k_y)$ using a generic sixth-order polynomial. **j-m,** Theoretical BDW and QM of CoSn (l,m) and their experimental estimates using $\tilde{g}$ and $\tilde{g}/\Delta E$ (j,k). $\Delta E = E_{flat} - E_{Dirac}$ is the energy difference between the flat band and the lower Dirac band. Theoretical quasi-BDW and QM are obtained from the $\{d_{xz}, d_{yz}\}$ orbital-based kagome tight-binding model developed to describe the CoSn dispersion.

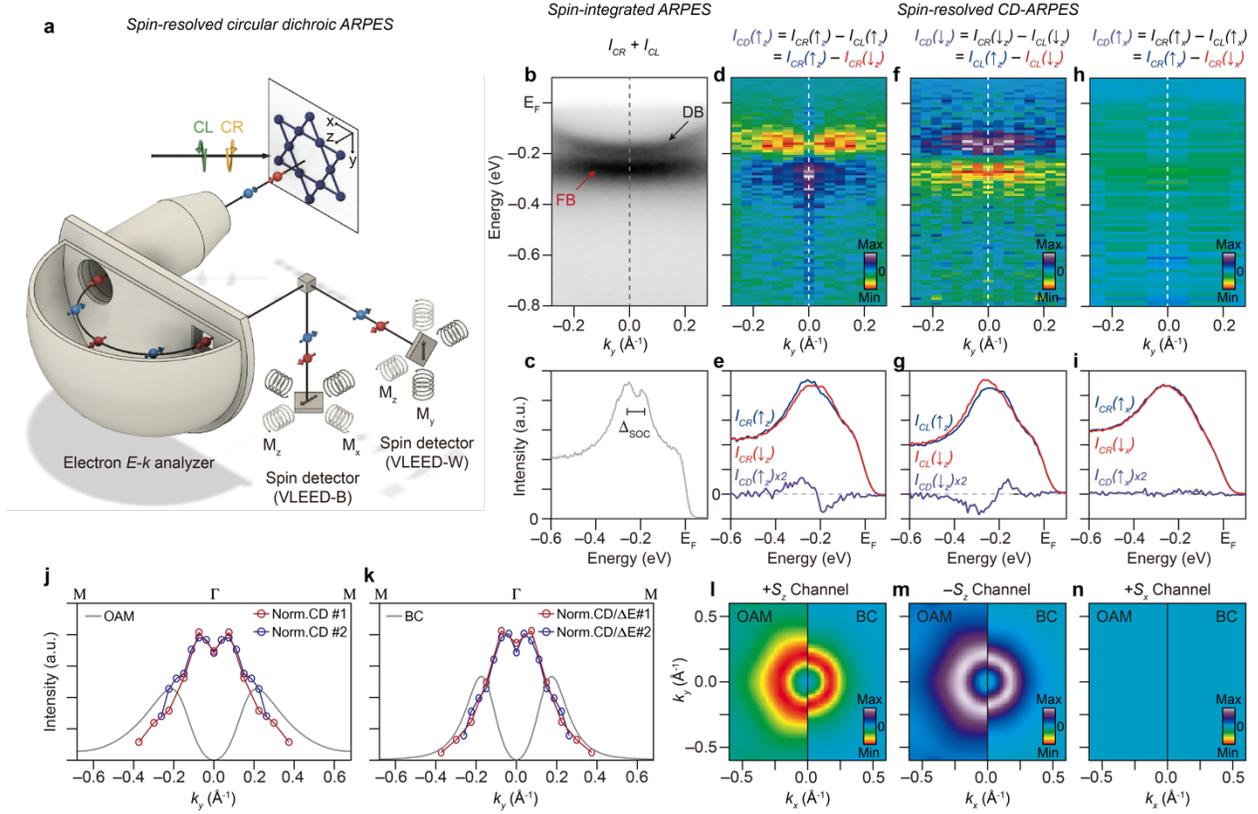

**Figure 4 | Experimental estimates of the OAM and Berry curvature. a,** Schematics of the spin-resolved CD-ARPES setup. The double VLEED spin detectors allow us to resolve spin components in full three-dimensional spin Hilbert space, the in-plane $S_x$, $S_y$, and the out-of-plane $S_z$. **b,c,** Spin-integrated ARPES spectra (b) and energy distribution curve at Γ (c) of CoSn. The $d_{xz}$ Dirac band, flat band, and SOC gap in between are marked in the respective panels. **d-i,** Spin-resolved CD-ARPES spectra and energy distribution curves at Γ measured in $+S_z$ (d,e), $-S_z$ (f,g), and $+S_x$ (h,i) spin channels, respectively. To avoid an artifact from the polarization-dependent beam intensities, we obtained $I_{CD}(\uparrow_z)$ as the difference between $I_{CR}(\uparrow_z)$ and $I_{CR}(\downarrow_z)$, using the fact that $I_{CR}(\downarrow_z)$ is identical to the $I_{CL}(\uparrow_z)$ by the time-reversal symmetry (see panel e). The same strategy has been applied to CD in other spin channels (panels g,i). **l-n,** Theoretical OAM and BC of the $d_{xz}$ Dirac bands in $+S_z$, $-S_z$, and $+S_x$ spin channels obtained from the $\{d_{xz}, d_{yz}\}$ orbital-based kagome tight-binding model of CoSn. **j,** Comparison between $I_{CD}^{norm}(\uparrow_z)$ and OAM along the Γ-M momentum-space direction. **k,** Corresponding comparison between BC and $-2I_{CD}^{norm}(\uparrow_z)/(E_{flat} - E_{Dirac})$. The experimental data in j,k are obtained from the two different CoSn samples (#1 and #2).